\documentclass[a4paper,fleqn,usenatbib]{mnras}

\usepackage{newtxtext,newtxmath}

\usepackage[T1]{fontenc}
\usepackage{ae,aecompl}

\usepackage{graphicx}	
\usepackage{amsmath}	
\usepackage{amssymb}	

\title[Ion and electron heating at the bow shock]{Stochastic ion and electron heating on drift instabilities at the bow shock}

\author[K. Stasiewicz]{
Krzysztof Stasiewicz,$^{1,2}$\thanks{E-mail: krzy.stasiewicz@gmail.com}
\\
$^{1}$Department of Physics and Astronomy, University of Zielona G\'ora, Poland\\
$^{2}$Space Research Centre, Polish Academy of Sciences, Warsaw, Poland\\
}

\date{Accepted 2020 May 11. Received 2020 April 01; in original form 2020 March 05 }

\pubyear{2020}

\begin{document}
\label{firstpage}
\pagerange{\pageref{firstpage}--\pageref{lastpage}}
\maketitle

\begin{abstract}
The analysis of the wave content inside a perpendicular bow shock indicates that  heating of ions is related to the Lower-Hybrid-Drift (LHD) instability, and heating of electrons to the Electron-Cyclotron-Drift (ECD) instability. Both processes represent stochastic acceleration caused by the electric field gradients on the electron gyroradius scales, produced by the two instabilities. Stochastic heating is a single particle mechanism where large  gradients break adiabatic invariants and expose particles to direct acceleration by the DC- and wave-fields.  The acceleration is controlled by function $\chi = m_iq_i^{-1} B^{-2}$div(\textbf{E}), which represents a general diagnostic tool for processes of energy transfer between  electromagnetic fields and particles, and  the measure of the local charge non-neutrality.  The identification was made with multipoint measurements obtained from the  Magnetospheric Multiscale spacecraft (MMS).  The source for the LHD instability is the diamagnetic drift of ions, and for the ECD instability the source  is ExB drift of electrons. The conclusions are  supported  by laboratory diagnostics of the ECD instability in Hall ion thrusters.
\end{abstract}




\section{Introduction}
Terrestrial bow shock represents a great opportunity for investigation of various mechanisms for heating and acceleration of particles in collisionless plasmas with important implications for astrophysics. Since its discovery by \citet{Ness:1964} there has been great deal of  research on this collisionless shock wave \citep{Wu:1984,Gary:1993,Balikhin:1994,Lembege:2003,Lefebrve:2007,Treumann:2009,Burgess:2012,Mozer:2013,Breneman:2013,Wilson:2014,Cohen:2019}, however, there is still no consensus on how particles are accelerated, and what exactly are processes that heat bulk plasma. 

The observational advances afforded by multipoint measurements in space  like Cluster \citep{Escoubet:1997}, THEMIS \citep{Sibeck:2008}, and 
MMS \citep{Burch:2016} opened new possibilities for space plasma physics. 
In this report we provide arguments supported by measurements from  the MMS mission, that  a single, non-resonant, frequency independent mechanism can mediate heating of both ions and electrons at the bow shock. This mechanism relies on gradients of the electric field that ensure breaking of the magnetic moment, allowing for efficient, stochastic heating of particles by the present electric fluctuations, and even by the DC field.

The energisation process is controlled by a dimensionless function defined as
\begin{equation}
\chi (t,\mathbf{r})  = \frac{ m_i}{q_i B^2} \nabla\cdot \mathbf{E}    \equiv \frac{N_c}{N}\frac{c^2}{V_{Ai}^2}  
\end{equation}
for particles with mass $m_i$, charge $q_i$  in fields $\mathbf{B}(t,\mathbf{r})$ and $\mathbf{E}(t,\mathbf{r})$. $N_c$ is the number density of excess charges, $N$ number density, $V_{Ai}^2=B^2/(\mu_0 N m_i)$, $c$ - speed of light. The equivalent formula is obtained with substitution $\nabla\cdot \mathbf{E}=N_cq_i/\epsilon_0$. Stochastic heating occurs when $|\chi |>1$. It is a single particle mechanism where large electric field gradients (or space charges) destabilise individual particle motion, rendering the trajectories chaotic in the sense of a positive Lyapunov exponent for  initially nearby states.  This condition has been known from some time, but the importance of div(\textbf{E}) has not  been recognised, and the directional  gradient $\partial_xE_x$, or  $k_\perp\nabla \phi $ was used in previous analyses or simulations \citep{Cole:1976,McChesney:1987,Karney:1979,Balikhin:1993,Mishin:1998,Stasiewicz:2000,Stasiewicz:2007,Vranjes:2010,Stasiewicz:2013,Yoon:2019}.

 The recent MMS mission comprises four spacecraft flying in formation with spacing $\sim$ 20 km and provides high quality 3-axis measurements of the electric field \citep{Lindqvist:2016,Ergun:2016,Torbert:2016}, which we shall use to demonstrate applicability of (1)  to heating of plasma at the bow shock.  We use magnetic field vectors measured by the Fluxgate Magnetometer \citep{Russell:2016}, the Search-Coil Magnetometer \citep{LeContel:2016},  and the number density, velocity, and temperature of both ions and electrons from the Fast Plasma Investigation  \citep{Pollock:2016}.

\section{Observations}
 We show in Fig.~1 the time-frequency spectrogram of $\chi$ for the  bow shock encountered by MMS  on 2017-12-12T03:33:30 at position (8.9, 11.8, 4.9) R$_E$ GSE. The shock had alfv\'enic and sound Mach numbers $M_A$=8, and $M_S$=5, and was strictly perpendicular. 
 The timeseries for $\chi$ was derived from 4--point measurements using a general method for computing gradients in space developed for Cluster \citep{Harvey:1998}. Frequencies lower than 1 Hz have been removed from the analysis to ensure that calibration offsets, or satellite spin effects  (0.05 Hz, and multiples) would not affect computations.

 The spectrum of $\chi$  is  similar to  the usual  power spectrum of the electric field, but weighted with $B^{-2}$ so it emphasises the heating region in the foot/ramp of the shock, instead of the peak, where the power of $E$ maximises, but there is no heating.   
  It is sensitive to local charges related to electrostatic  fields generated in plasma by waves and instabilities.
 Ideally, it should show distribution of electric charges,  and time-domain variations at long wavelengths should vanish by subtraction.  However, the spacecraft separation of 10--20 km does not make it possible to compute correctly gradients on scales $\sim$ 1 km, which exist in this region. 

\begin{figure}
\includegraphics[width=\columnwidth]{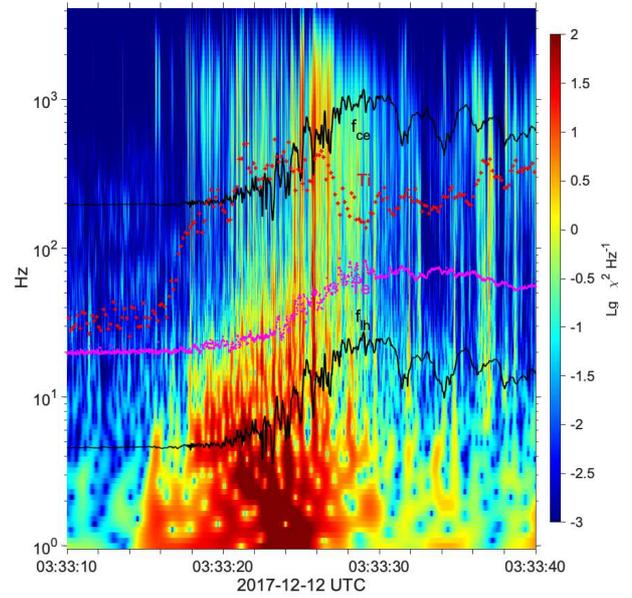}
\caption{Power spectrum of $\chi$ across the perpendicular bow shock. Over-plotted are: the lower-hybrid frequency $f_{lh}$ (black), temperatures,  $T_{e\perp}$ (magenta),  $T_{i\perp}$ (red), both in eV, and  electron cyclotron frequency $f_{ce}$.  At the bottom up to $f_{lh}$ there are oblique whistler waves, followed by striations of Lower-Hybrid-Drift waves in the range $f_{lh}-f_{ce}$ that extend as  Electron-Cyclotron-Drift waves above $f_{ce}$. \label{chi_spec}}
\end{figure}

We also show over-plots of the ion perpendicular temperature, electron temperature, (both in eV), and the lower-hybrid frequency $f_{lh}=f_{pp}/(1+f_{pe}^2/f_{ce}^2)^{-1/2}$, where $f_{pp}, f_{pe}$ are proton, electron plasma frequencies, $f_{ce}$ electron gyrofrequency.  The $f_{lh},\;f_{ce}$ plots represent  the density and magnetic field variations across the shock, which should help the reader to position the foot and the ramp of the shock. The proton gyrofrequency $f_{cp}$= 0.1--0.4 Hz, is just below the bottom line of the picture.

The power of $\chi$ correlates well with  regions of ion and electron heating, which is not true for the power spectrum of the electric field. A well known fact is that heating of ions is not co-located with heating of electrons, as seen also in the picture.
Evident in the picture is a turbulent cascade that appears to transfer energy from  lower frequency modes at the bottom to higher frequencies  in localised striations extending to 4096 Hz, the upper frequency of measurements. Waves above $f_{lh}$ are identified further as LHD and above $f_{ce}$ as ECD.

We now turn to a different technique of signal analysis -- namely, multiresolution frequency decomposition using orthogonal wavelets \citep{Mallat:1999}. 
 This technique differs from pass-band filtering.  Orthogonal decomposition is exact; the signal is divided into discrete frequency dyads that form $2^{-n}f_N$ hierarchy starting from the Nyquist frequency $f_N$.  Orthogonality means that time integral of any pair of the frequency dyads is zero. The decomposed electric field is shown in Fig.~2 grouped into three layers with perpendicular and parallel components shown separately. In order to save space, plots show only  halves of the waveforms for the parallel and for one  of the perpendicular components. Similar decomposition for the magnetic field is shown in Fig.~3.

Noticeable in these plots is the electrostatic character of waves in panel 2a and electromagnetic character of  waves in panels 2c, 3c. Also interesting is the large parallel electric field, as well as the compressional magnetic field present in all wave modes.

\begin{figure}
\includegraphics[width=\columnwidth]{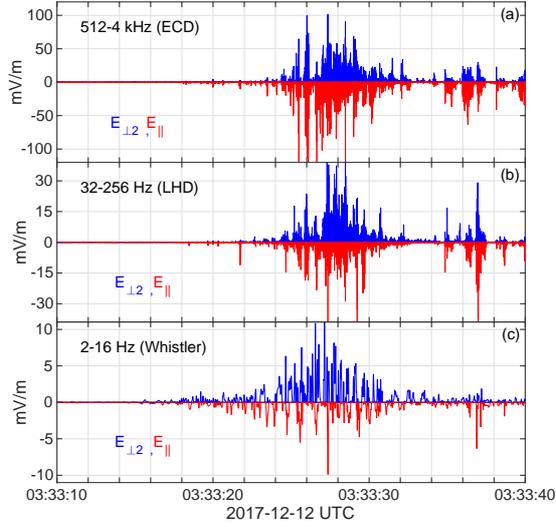}
\caption{The measured electric vector, sampled at 8192 s$^{-1}$ is  decomposed into  3 bands using orthogonal wavelets (complete signal is exact sum of the components) and     transformed to magnetic field-aligned coordinates (FAC). From the bottom: (c) obliquely propagating whistler waves (2--16 Hz),  (b) Lower-Hybrid-Drift waves (32--256 Hz), and (a) Electron-Cyclotron-Drift waves (512--4000 Hz). To  enhance readability we show only halves of the waveforms for the parallel and for one of the perpendicular components. \label{e_dec}}
\end{figure}

\begin{figure}
\includegraphics[width=\columnwidth]{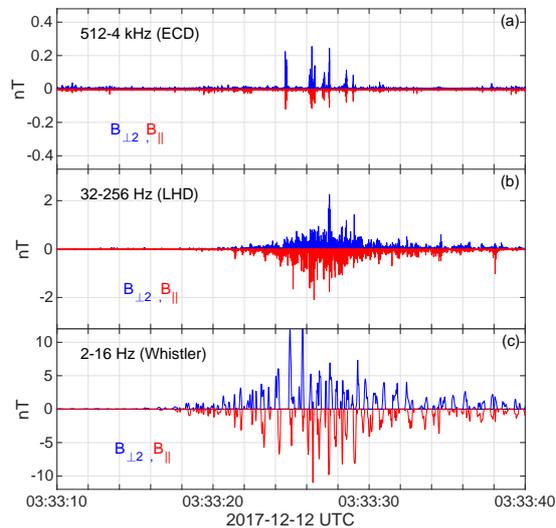}
\caption{The measured magnetic signal is decomposed into  3 bands  and shown in FAC coordinates as in Fig.~2.  \label{b_dec}}
\end{figure}

Frequency range between $f_{cp}$ and $f_{ce}$ is occupied by whistler  and lower-hybrid waves which belong to the same branch of the dispersion equation but differ by the propagation direction and polarisation properties.  LH waves propagate close to perpendicular direction and are mostly electrostatic, while whistlers have k-vectors in a wide angular range and become purely  electromagnetic in the parallel direction. These two wave types can change their mode by conversion on the density gradients and striations \citep{Rosenberg:2001,Eliasson:2008,Camporeale:2012} that exist at the bow shock. 

The analysis of the propagation direction for individual frequency dyads in the range 2--16 Hz shows that they propagate 120--140 degrees to the magnetic field, which implies  that these are oblique whistler modes. The propagation direction is established both from the Poynting vector direction and the minimum variance of the wave magnetic field.

At 32 Hz and above, the propagation direction becomes perpendicular,  the magnetic component diminishes rapidly with increasing frequency, indicating electrostatic, lower-hybrid character. These waves are most likely related to the LHD instability, which is a cross-field current-driven instability that couples LH waves with drift waves generated on the density gradients \citep{Krall:1971,Davidson:1977,Daughton:2003}.   According to theory, there could be two types of instability. A weaker one -- kinetic, should grow when   the  scale of the density gradient $L_n= (N^{-1}|\nabla N|)^{-1}$ is in the range $ 1<L_n/r_p <(m_p/m_e)^{1/4}$, while the stronger, fluid-type instability occurs when $L_n/r_p<1$.  Here, $r_p$ is a proton thermal gyroradius. The instability is caused by the  diamagnetic ion drift $V_{d}=T_p(m_p\omega_{cp}L_n)^{-1}=v_{tp}(r_p/L_n)$. The maximum growth rate is at $k_\perp r_e\sim$1, i.e. at wavelengths of a few electron gyroradii. These wavelengths will be Doppler shifted by the prevailing plasma flows $V\approx$300 km/s to frequencies $f=V/(2\pi\lambda)\sim$ 40-70 Hz, in our case. 

The scale $L_n$ of the density gradient at the bow shock that can be determined directly by the MMS is indeed smaller than the ion gyroradius, as seen in Fig.~4a.  This provides additional support for the interpretation that the bursty  waves in  the range $f_{lh}-f_{ce}$ of Figures 1 and 2b are mostly LHD waves. They are probably a permanent feature of quasi-perpendicular bow shocks because of the density ramp that drives the  instability. Such waves have been observed also in other regions of the magnetosphere \citep{Bale:2002,Vaivads:2004,Norgren:2012,Graham:2017}.

\begin{figure}
\includegraphics[width=\columnwidth]{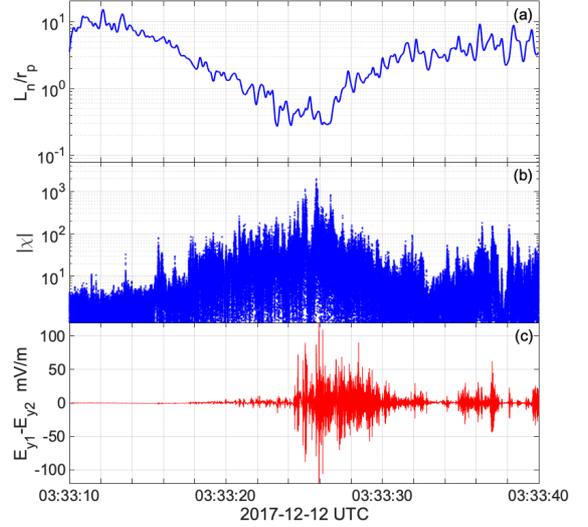}
\caption{(a) Computed gradient scale of the density $L_n$ normalised with proton gyroradius $r_p$ showing that the region is strongly unstable for Lower-Hybrid-Drift waves ($L_n/r_p<$1). (b) Timeseries of $|\chi |$, which correlates well with heating of ions and electrons. It contains 245,760 data points. (c) Difference $E_{y1}-E_{y2}$ of the electric field measured by satellites 1 and 2, separated by $\Delta y$=5 km.\label{Lchi}}
\end{figure}

Panel 2a contains  electrostatic waves, which have been classified by first observers \citep{Rodriguez:1975,Fuselier:1984} as ion-acoustic modes because of the large parallel electric field component. Waves  in this frequency range have been analyzed with use of high-time resolution measurements obtained by  THEMIS \citep{Mozer:2013,Wilson:2014}, and by STEREO and Wind \citep{Breneman:2013}, who noted large parallel electric fields and identified  electron cyclotron harmonics in the spectra. Their conclusion that these are ECD waves is supported by the spectrum in Fig.~5  which shows seven $f_{ce}$ harmonics.

\begin{figure}
\includegraphics[width=\columnwidth]{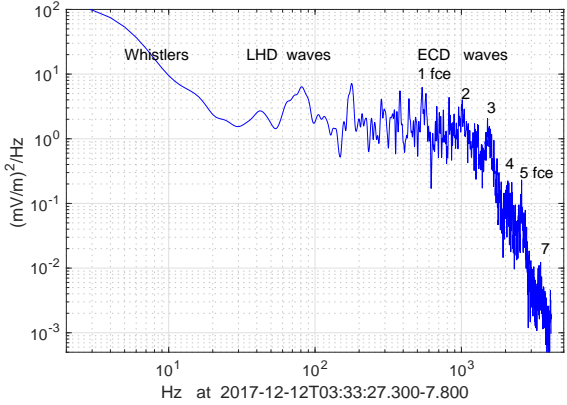}
\caption{Power spectrum of $E_\perp$ from the electron heating region. Obliquely propagating whistlers extend from 2 up to $f_{lh}\sim$16 Hz followed by LHD waves. Clearly seen are Electron-Cyclotron-Drift waves with 7 harmonics starting from $f_{ce}\approx$500 Hz.  \label{fspectrum}}
\end{figure}

The  ECD instability \citep{Forslund:1972} is caused by the perpendicular relative ion-electron drift $V_d$ comparable to the thermal electron speed $v_{te}$. It occurs at the resonance $k(V_d-v_{tp})=n\omega_{ce}$, couples electron Bernstein modes with ion-acoustic waves, and produces wavelengths smaller than the electron gyroradius \citep{Lashmore:1971}. 
 In the current theoretical scenario \citep{Muschietti:2013} this instability is presumably produced by ion beams reflected from the shock and moving perpendicularly against the solar wind and stationary electrons. This is a rather unrealistic scenario because two ion populations cannot have perpendicular ExB drift in opposite directions at the same time, and different from electrons (there are no other drifts in their model). This model would put the location of ion-beam driven ECD instability in front of the shock, around time 03:33:15 in our data, while the maximum of ECD waves is at 03:33:27 in the electron heating region.  Despite the unrealistic scenario,  the mathematics and conclusions of this paper are correct and valuable. 

The data presented here suggest another mechanism for the ECD instability. The LHD electric fields of 20 mV/m  in $B\sim$15 nT produce ExB drift  of 1300 km/s, comparable to the electron thermal speed. Since this field exists in narrow channels with a width of a few electron gyro-radii, only electrons can participate, making ion-electron drift $V_d\sim v_{te}$.
Incidentally, this instability attracted attention of researchers working with Hall ion thrusters, because it affects their performance  \citep{Ducrocq:2006,Boeuf:2018}. In these devices, the instability  is also caused by the ExB drift of electrons only, because ions have gyroradius larger than the drift channels.

The theory of ECD instability that links electron cyclotron harmonics with ion-acoustic waves and wavelengths at scales below the electron gyroradius provides  convincing explanation for waves in panel 2a, which exhibit both large $ E_\parallel$ and $E_\perp$, contain $f_{ce}$ harmonics, and Doppler shifted  structures smaller than electron gyroradius.

\section{What heats ions and electrons?}

We focus now on the fundamental question what kind of processes can increase the ion energy by 400 eV (from 50 to 450), and the electron energy  by 50 eV (from 20 to 70 eV),  within a few seconds, or during one proton gyro-period, as seen in Fig.~1.  Additional constraint is that heating of electrons  is isotropic, while ions is mostly perpendicular.  

The mechanism of stochastic ion heating on LH waves related to condition (1) has been explained by \citet{Karney:1979}, and on other wave structures by other authors: \citet{Cole:1976,McChesney:1987,Balikhin:1993,Mishin:1998,Stasiewicz:2000,Stasiewicz:2007,Vranjes:2010,Stasiewicz:2013,Yoon:2019}. The essence is that particles stressed by strong gradients (or equivalently by local space charges, div(\textbf{E})=$\rho/\epsilon_0$) loose adiabaticity and can be scattered along the electric field for direct acceleration by both  DC- and wave-fields. 
 The computed $\chi$ (Fig.~4b) shows values greatly exceeding 1, needed to stochastically perturb proton orbits, however, significant ion heating is observed when $\chi>10$.

 Formally, the error of $\chi$ is equivalent to the error of the electric field measurements, i.e., $\sim$1 mV/m, or 10\% of large amplitude fields, because of gain uncertainty in amplifiers \citep{Lindqvist:2016}.  However, in the case of  ECD waves with wavelengths of the electron gyroradius, the gradient computations underestimate real values because the spacecraft separations are larger than wavelengths. 
  In Figure 4c we show the difference in $E_y$ components measured by satellites 1 and 2. Similar differences are seen on any pair of the satellites. A difference $\Delta E_y\approx$100 mV/m on scales of the electron gyroradius $r_e\approx$1 km in the field $B$=15 nT would produce  $\chi\approx$4600, which could effectively demagnetise electrons and subject them to stochastic heating. Applying 10\% error to this value would still maintain $\chi$ well above the required threshold for electron heating, which is reached in computations shown in Fig. 4b.

 The presented data imply the following scenario  for the rapid ion heating seen in Fig.~1. The diamagnetic current related to increasing density gradient  at the foot of the shock (Fig.~4a) drives  LH waves that couple/convert to oblique whistlers on density gradients and turn into LHD instability when the threshold is exceeded ($V_d\sim v_{tp}$). This results in increase of the wave amplitude, and cascade to shorter scales ($kr_e\sim$1), which implies  $\chi>10$ that can perturb incoming solar wind ions. The ion energisation  is probably a one step process \citep{Stasiewicz:2007} (see Fig.~1 in that paper) in which ions gain energy from the DC electric field and not from wave fields. There is a quasi-DC (below 1 Hz) electric field of $E_0$=5 mV/m in this region. Protons stochastically perturbed by $\chi\sim$10--100 on LHD waves (Fig.~4b),  can be scattered along $\mathbf{E}_0$, and acquire  large perpendicular velocity during one gyro-period.  To acquire 400 eV required for ion heating one needs  electric potential from $E_0$=5 mV/m on a distance of 80 km. A displacement of   ion gyrocenters by less than a gyroradius is sufficient to account for ion energisation. 
 
Electrons require $\chi$ larger by a factor $m_p/m_e$ to start stochastic heating. Apparently, amplitudes of LHD waves cannot produce gradients strong enough to make $\chi>$1836 in the ion heating region of Fig.~1. However, as mentioned earlier, the electric field of LHD waves at $\sim$20 mV/m can produce ExB drift of electrons  comparable to the thermal velocity ($V_d=V_E\sim v_{te}$), which is needed to start ECD instability. This instability will produce stronger fields $\sim$150 mV  (Fig.~2a) and cascade to scales less than electron gyroradius, which would make $\chi>1836$ and initiate electron heating. The mechanism for the ECD instability at the bow shock is similar to that occurring in Hall thrusters. In both cases only electrons are drifting, because ions are prevented from the ExB drift by large gyroradius compared to the size of drift channels.

Stochastic heating of electrons occurs  in  localised bursts  and is quenched by increasing $B$, because of the dependence   $V_E\propto B^{-1}$ for the instability driver, and  $\chi\propto B^{-2}$ for the stochastic condition.
In most of the region, $\chi<m_p/m_e$, so the electrons respond adiabatically, $T_{e\perp} \propto B$, and a major part of the energy increase for  electrons can be attributed to the increasing $B$ at the ramp of the shock. Such perpendicular adiabatic heating must be accompanied by isotropization, possibly by $E_\parallel$ of waves shown in  Figure 2a.

It is interesting to note that by expressing div(\textbf{E}) $\sim E/L_E$, and $V_E =E/B$ we can rewrite the condition $\chi_e>1$ as
\begin{equation}
\frac{V_E}{v_{te}}>\frac{L_E}{r_e}
\end{equation}
which implies that to start electron heating  we need the ratio of the drift to thermal velocity exceeding the ratio of the electric gradient scale to the gyroradius. This fits  properties of ECD waves, and the instability condition. 

Another implication of equation (1) is the connection to charge non-neutrality. With typical $V_A\sim$ 100 km/s at the bow shock, we need local charge non-neutrality exceeding 10$^{-7}$ to start ion heating, and  2$\times$10$^{-4}$ to heat electrons. Also the presence of $E_\parallel\sim E_\perp$ in ECD waves in Fig.~2a is a natural consequence of space charges, which would produce  electric fields in both directions.

\section{Conclusions}

The observations and the above analysis support the  hypothesis on the universal character of the relation (1) and its applicability for identifying the heating processes of both electrons and ions, independent of the wave mode and the type of  instability.
 
The plasma heating mechanism at quasi-perpendicular bow shocks appears to be a two-stage process. In the first stage, the {\em ion diamagnetic drift}  on the density gradients, $V_d \sim v_{tp}$, ignites the LHD instability that produces $E\sim$20 mV/m, and $\chi\gg1$ that heats ions. 
In the second stage, the  electric field of LHD waves produces {\em electron ExB drift},  $V_E\sim v_{te}$, which ignites the ECD instability that produces  stronger fields, $E\sim$100 mV/m, and $\chi>m_p/m_e$, that heats electrons. Part of the electron energy increase can be attributed to adiabatic heating $T_{e\perp} \propto B$ with isotropization by ECD waves.

The mechanism of the ECD instability at the bow shock is similar to the experimentally verified ECD instability in ion Hall thrusters, i.e., the ExB electron drift. All elements of the heating scenario described above have  experimental support in MMS measurements.

Finally, it should be emphasised that this paper presents for the first time spectrum of the divergence of the electric field measured in space.  Figure 1 is a time-frequency spectrogram of the electric charge distribution, weighted with $B^{-2}$,  inside the bow shock. Obviously, it is an approximation, because gradients on small scales are not well resolved. The credits should go entirely to the members of the FIELDS consortium of the MMS project   \citep{Torbert:2016,Lindqvist:2016,Ergun:2016} for designing the instruments and performing the project.

\section*{Acknowledgements}
Special thanks to members of the FIELDS consortium  of the MMS mission for making instruments capable of measuring divergence of the electric field in space. The author is grateful to Yuri Khotyaintsev for his invaluable help with the software for data processing. The MMS data are available to public via  https://lasp.colorado.edu/mms/sdc/public/





\bibliographystyle{mnras}



\bsp	
\label{lastpage}
\end{document}